\documentclass[preprint,twocolumn,10pt,a4paper,byrevtex,prb,unsortedaddress,superscriptaddress]{revtex4}

\usepackage{amssymb}
\usepackage{soul}
\usepackage{graphicx}
\usepackage{hyperref}
\usepackage{subfigure}
\usepackage{amsmath}
\usepackage{color}
\usepackage{natbib}
\usepackage{appendix}

\definecolor{DarkGreen}{rgb}{0,0,0}
\newcommand{\farkhad}[1]{\textcolor{DarkGreen}{#1}}

\begin{document}

\title{Boosting room temperature tunnel magnetoresistance in hybrid magnetic tunnel junctions under electric bias}

\author{C\'esar Gonz\'alez-Ruano}
\affiliation{Departamento F\'isica de la Materia Condensada C-III, Instituto Nicol\'as Cabrera (INC) and  Condensed Matter Physics Institute (IFIMAC), Universidad Aut\'onoma de Madrid, Madrid 28049, Spain}

\author{Coriolan Tiusan}
\affiliation{Department of Physics and Chemistry, Center of Superconductivity Spintronics and Surface Science C4S, Technical University of Cluj-Napoca, Cluj-Napoca, 400114, Romania}
\affiliation{Institut Jean Lamour, Nancy Universit\`{e}, 54506 Vandoeuvre-les-Nancy Cedex, France}

\author{Michel Hehn}
\affiliation{Institut Jean Lamour, Nancy Universit\`{e}, 54506 Vandoeuvre-les-Nancy Cedex, France}

\author{Farkhad G. Aliev}
\email[e-mail: ]{farkhad.aliev@uam.es}
\affiliation{Departamento F\'isica de la Materia Condensada C-III, Instituto Nicol\'as Cabrera (INC) and  Condensed Matter Physics Institute (IFIMAC), Universidad Aut\'onoma de Madrid, Madrid 28049, Spain}
%\date{\today }

\begin{abstract}

Spin-resolved electron symmetry filtering is a key mechanism behind giant tunneling magnetoresistance (TMR) in Fe/MgO/Fe and similar magnetic tunnel junctions (MTJs), providing room temperature functionality in modern spin electronics. However, the core process of the electron symmetry filtering breaks down under applied bias, dramatically reducing the TMR above 0.5 V. This strongly hampers the application range of MTJs. To circumvent the problem, resonant tunneling between ferromagnetic electrodes through quantum well states in thin layers has been used so far. This mechanism, however, is mainly effective at low temperatures. Here, a fundamentally different approach is demonstrated, providing a strong TMR boost under applied bias in V/MgO/Fe/MgO/Fe/Co hybrids. This pathway uses spin orbit coupling (SOC) controlled interfacial states in vanadium, which contrary to the V(001) bulk states are allowed to tunnel to Fe(001) at low biases. The experimentally observed strong increase of TMR with bias is modelled using two nonlinear resistances in series, with the low bias conductance of the first (V/MgO/Fe) element being boosted by the SOC-controlled interfacial states, while the conductance of the second (Fe/MgO/Fe) junctions controlled by the relative alignment of the two ferromagnetic layers. These results pave a way to unexplored and fundamentally different spintronic device schemes, with tunneling magnetoresistance uplifted under applied electric bias.

\end{abstract}

\maketitle

\section{Introduction}

Nearly two decades after atomistic calculations predicted the giant TMR due to symmetry resolved coherent tunneling \cite{Butler2001,Mathon2001}, this effect is currently one of the basic electron transport mechanisms in most of spintronics sub-branches, ranging from magnetoresistive random access memories and programmable logic elements \cite{Ney2003} to magnetic sensors \cite{Lenz2006}. The theoretically predicted several-thousand-percent zero bias TMR is, in practice, reduced an order of magnitude due to interface inter-diffusion, roughness or surface states.

However, even more humbling basic characteristic of single barrier MTJs is a \textit{strong monotonic reduction of the TMR with voltage}, on the scale of about 0.5 V \cite{Parkin2004,Yuasa2004,Waldron2006,Guerrero2007,Herranz2010}. \farkhad{For the case of incoherent tunneling in non-epitaxial junctions, the reduction is mainly caused by itinerant tunnel electrons with excess energy above the Fermi level ($E_F$) (the so-called “hot electrons”) \cite{Parkin1997}. These electrons, which are generated with increasing bias and temperature, are able to excite magnons at the FM/I interface, which in turn favour spin-flip events in the tunneling process \cite{Moodera1995}. While this is not relevant for a spin valve in the parallel (P) configuration (since no change in spin is necessary for the transport), it is significant in the anti-parallel (AP) configuration, increasing the transmission and thus reducing the resistance difference with the P state. For the coherent tunneling case, as in our epitaxial junctions, the scenario involves the band-to-band transmission of electrons preserving the orbital symmetry \cite{Guerrero2007}. The applied bias increases the coupling between the two FM electrodes, creating new channels that enhance the transmission in the AP state faster than in the P state \cite{Waldron2006}.}

So far, a few different approaches have been used to compensate the main shortcoming of MTJs for the applications: (i) optimising the ferromagnetic electrodes (F1,F2) or barrier (I) in F1/I/F2 MTJs, (ii) enhancing the conductance for specific biases through resonant tunneling via quantum well states (QWS) inside the barrier in F1/QWS/I/F2 MTJs, or (iii) sequential tunneling in F1/I/F2/I/F3 double barrier MTJs (DMTJs).

Within the first approach (i), asymmetric single barrier junctions have been proposed in order to partially mitigate the electric-field-induced TMR degradation, suppressing the conductance through surface states for some bias polarity \cite{Zhang2004}. On the other hand, the incorporation of EuS spin-filter barriers \cite{Nagahama2007} is only operative below the EuS Curie temperature ($T_C=17$ K). 

Spin-dependent resonant tunneling through quantum-well \cite{Lu2005,PeraltaRamos2008} and interface \cite{Zhang2004} states investigated within line (ii) showed that devices are either operative at low temperatures only or involve technically challenging symmetry confinement in a few mono-layers of ferromagnetic material \cite{Xiang2019}. Other similar \cite{Niizeki2008} approaches providing TMR modulation are also not robust enough to build energy efficient \cite{Kowalska2019} room temperature spintronic devices. Finally, within the approach (iii) only a TMR suppression slower than the expected for two independent MTJs in series has been achieved in standard DMTJs \cite{Montaigne1998}, while showing a steady TMR decay with bias \cite{Tiusan2006,Gan2010}.

\begin{figure*}[ht]
\begin{center}
\includegraphics[width=1.8\columnwidth]{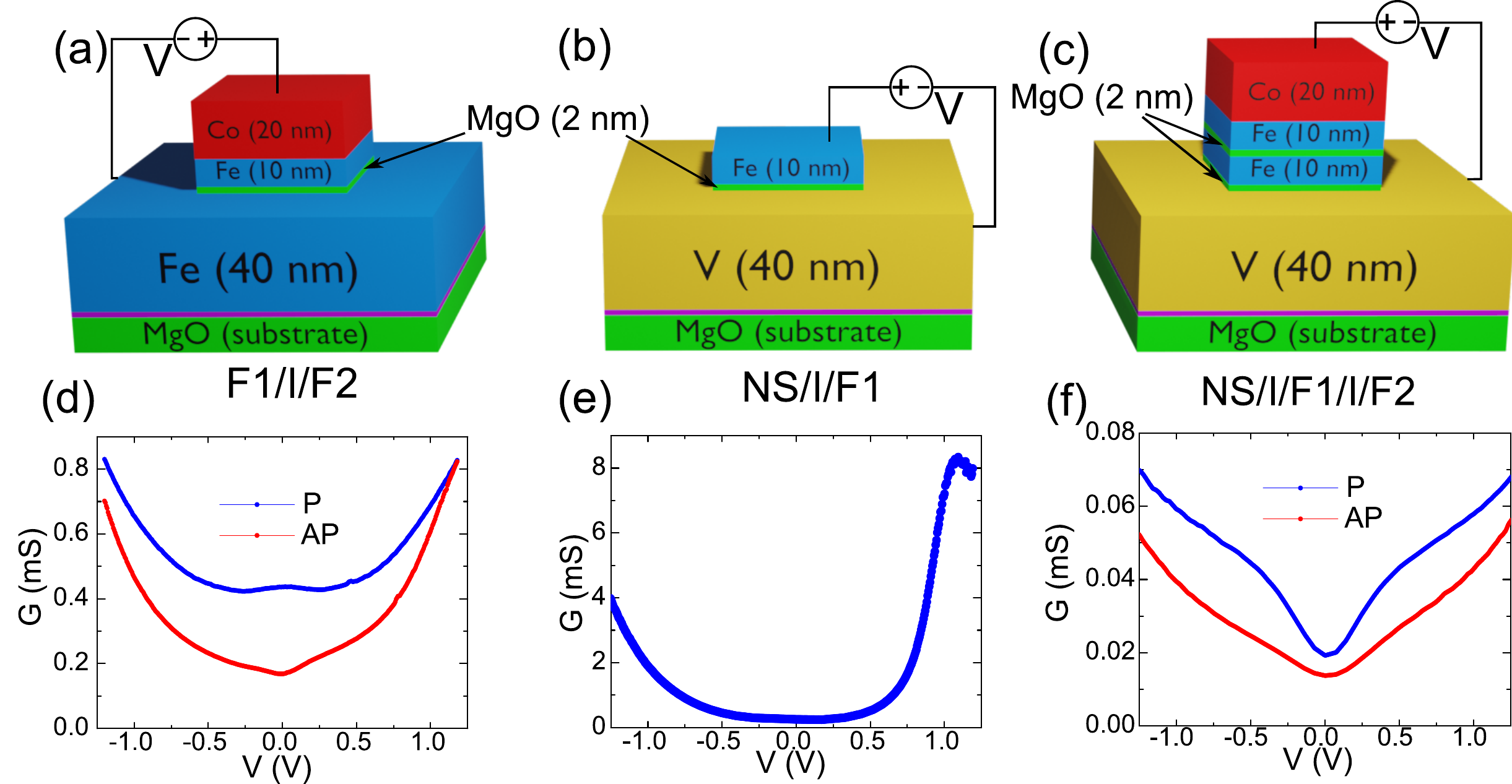}
\caption{(a), (b) and (c) show sketches of the F1/I/F2 (control samples), NS/I/F1 and NS/I/F1/I/F2 junctions under study, respectively. Vanadium (001) is the NS electrode, while F1 and F2 are the respectively magnetically soft Fe(001) and hard (sensing) Fe(10 nm)Co(20 nm) ferromagnetic layers. (d) Shows conductance curves in the P and AP configurations of a control sample at room temperature. Part (e) shows the high bias conductance in a V/MgO/Fe junction at room temperature, evidencing the conductance variation with voltage of the NS/I/F1 junctions, of nearly two orders of magnitude below 1 V. Part (f) shows the high bias conductance in a V/MgO/Fe/MgO/Fe/Co junction in the P (blue) and AP (red) alignments of the ferromagnetic electrodes, also at room temperature.}
\label{Fig1}
\end{center}
\end{figure*}

In this study we demonstrate a radically different and breakthrough approach to solve the long standing problem. 
We report an electric-bias-induced 2-3 fold increase of the TMR in hybrid MTJs (H-MTJs) of V/MgO/Fe/MgO/Fe/Co (NS/I/F1/I/F2 with NS being a normal metal with surface states) in a wide temperature range, from liquid helium to room temperatures. The \textit{output voltage} parameter, which is crucial for possible applications \cite{Gan2010,Tezuka2016,Useinov2012,Mukaiyama2016}, strongly exceeds the values observed for single barrier F1/I/F2 MTJs at biases above 0.5 V. Moreover, to our best knowledge, the room temperature value exceeding 0.8 V are a record high for all known spintronic devices. We explain this unprecedented behavior with a simplified model, considering magnetic-state-dependent sequential tunneling though two nonlinear devices in series. The conductance of the first element (V/MgO/Fe) is essentially controlled by SOC and symmetry protected surface states, while the conductance of the second (Fe/MgO/Fe/Co) is given by its magnetic state.

\section{Experimental results}

Details on the samples fabrication and measurement procedures are explained in the Methods section. {Figure \ref{Fig1}} introduces the experimental configurations for the F1/I/F2 control samples, the NS/I/F1 single barrier and the NS/I/F1/I/F2 double barrier hybrid structures. Panels a, b and c sketch the structure of each junction respectively, while panels d, e and f compare the bias dependencies of the conductance for the type of sample shown above. The lateral size of all the studied junctions is $20\times20~\mu\text{m}^2$ unless otherwise stated. The conductance of V/MgO/Fe shows a very strong bias dependence reaching about a $\times3$ increase up to 0.5 V and a $\times100$ up to 1.3 V.
Figure \ref{Fig1}f demonstrates that for the broad low bias range below about 0.5 V, the conductance of the NS/I/F1/I/F2 junction varies much more with bias in the P state compared to the AP state. This is clearly an opposite behavior with respect to the one observed in F1/I/F2 junctions (Figure \ref{Fig1}d).

\subsection{Bias enhanced TMR and dependence on temperature and junction size}

\begin{figure}[h]
\begin{center}
\includegraphics[width=\linewidth]{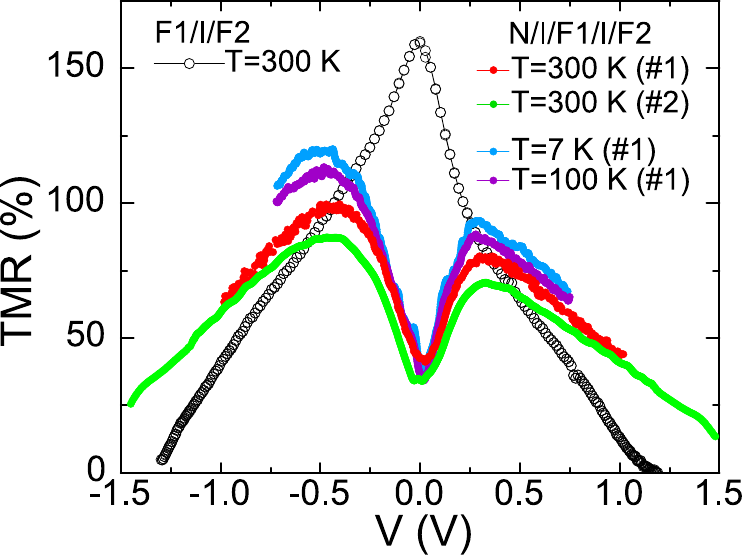}
\caption{Comparative of TMR vs applied bias for several different H-MTJs at different temperatures. Light blue ($T=7$ K), purple ($T=100$ K), red and green ($T=300 K$) curves correspond to two different N/I/F1/I/F2 junctions, labeled as \#1 and \#2 in the legend. The curve with black open dots corresponds to a Fe/MgO/Fe/Co control sample at $T=300$ K.}
\label{Fig2}
\end{center}
\end{figure}

{Figure \ref{Fig2}} compares the bias dependence of TMR for the single and double barrier spin valve structures.

The TMR is calculated as the relative conductance difference between the parallel (P) and antiparallel (AP) states of the two ferromagnetic layers:

\begin{equation}
\text{TMR}=\frac{G_{P}-G_{AP}}{G_{AP}}\times100
\end{equation}

For the Fe/MgO/Fe/Co single barrier structures (F1/I/F2), TMR is maximum near $V=0$ V, and decreases with increasing bias. On the contrary, for NS/I/F1/I/F2 H-MTJs the TMR first increases with voltage, until a maximum is reached near $V=0.5$ V, and then starts to decrease. This behavior is further enhanced at low temperatures, as can be seen in Figure \ref{Fig2}. 

\subsection{Bias dependence of output voltage}

\begin{figure}[h]
\begin{center}
\includegraphics[width=\linewidth]{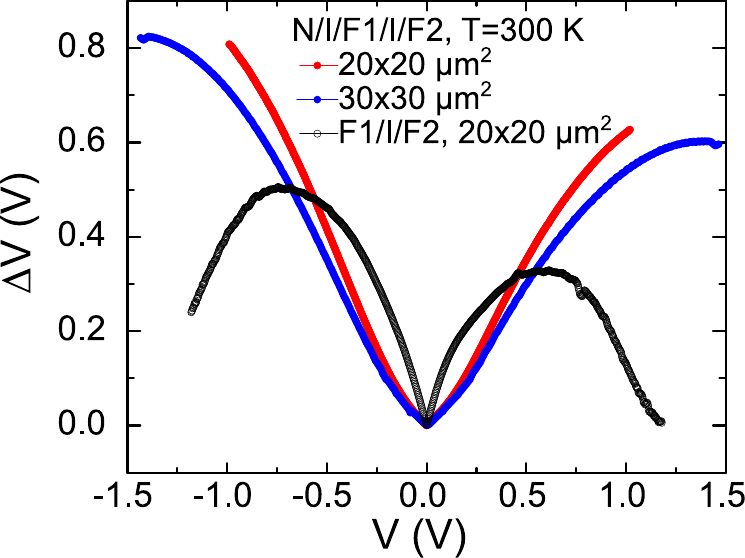}
\caption{$\Delta V$ vs applied bias at room temperature for a control sample (black) and two H-MTJs of different lateral sizes (red is a $20\times20$ $\mu\text{m}^2$ blue a $30\times30$ $\mu\text{m}^2$ one).}
\label{Fig3}
\end{center}
\end{figure}

Besides the TMR vs bias determined from the differential conductance, we have also analyzed a related parameter which is important for applications, namely the output voltage of the device \cite{Gan2010,Tezuka2016,Useinov2012,Mukaiyama2016}, defined as

\begin{equation}
\Delta V=\left|V\right|\times\frac{R_{AP}-R_P}{R_P}
\end{equation}

where the magnetoresistance is now defined from the resistance instead of the conductance. 
We observed that the output voltage $\Delta V$ in MTJs is higher than the one of H-MTJs only at low biases (see {Figure \ref{Fig3}}). For H-MTJ devices, the output voltage monotonously increases with voltage at room temperature practically in the whole measured range, exceeding 0.8 V at an applied bias of 1.5 V, while for the single barrier MTJs it peaks at $V\sim0.7$ V. The maximum $\Delta V$ reached exceeds previous record values in Fe/C/MgO/Fe tunnel junctions, which were also obtained with an applied bias of 1.5 V \cite{Tiusan2006_2}.

\section{Discussion} 

Below we discuss two most important properties of the V/MgO interface explaining the unexpected TMR vs bias behavior. Its strongly nonlinear conductance response to the applied bias is due to (i) extremely low zero bias conductance (as indicated by the experiments, see Figure \ref{Fig1}) and the presence of surface states with an electron symmetry fundamentally different from the bulk one.

\begin{figure*}[ht]
\begin{center}
\includegraphics[width=1.8\columnwidth]{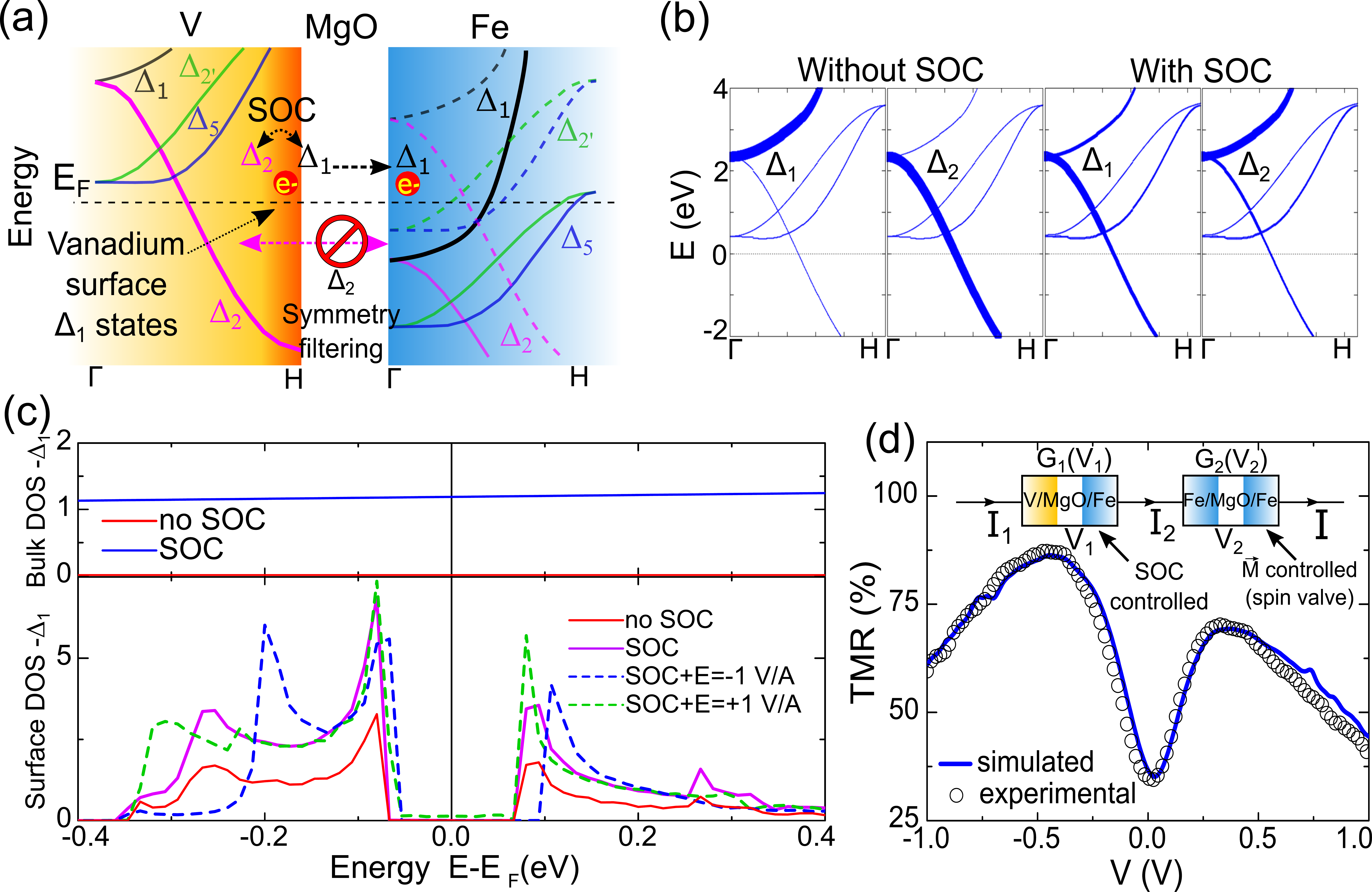}
\caption{(a) Sketch of the band structure of the V/MgO/Fe system, including the vanadium $\Delta_1$ surface states, representing the main transport mechanism involving interfacial SOC and symmetry filtering in the MgO. \farkhad{Similar results have been obtained in Ref. \cite{Feng2009}}. In the Fe part, solid lines indicate the bands for the main spin population, while dashed lines are used for the minority spin population bands. \farkhad{(b) Band structure of the V layer, in the absence of SOC (first 2 panels) and with SOC (third and fourth), where the thickness of the band line is proportional to the indicated electron symmetry. This is used to calculate the DOS in (c), which shows the} $\Delta_1$ density of states (DOS) in the Vanadium electrode, both in the bulk (upper part) and the surface (below), where new states appear around $\pm 0.1$ eV from the Fermi energy in the presence of SOC and electric field. (d) The TMR($V$) characteristic of the hybrid double barrier V/MgO/Fe/MgO/Fe/Co structure can be modelled as two in series nonlinear devices \farkhad{(sketch shown in inset)}. The system is solved considering the experimental $G_2(V_2)$ in P and AP configurations for the F1/F2 barrier and a parametrized $G_1(V_1)$ of the N/F1 barrier, adjusted to obtain the best fit of the final TMR($V$) curve for the N/F1/F2 junction.}
\label{Fig4}
\end{center}
\end{figure*}

Our ab-initio calculations ({Figure \ref{Fig4}}a-c), in line with previous studies \cite{Feng2009}, explain the very low zero bias conductance in V/MgO/Fe as only a single bulk band with $\Delta_2$ symmetry crosses the Fermi energy in the [100] direction (normal to the interface). The $\Delta_2$ symmetry states, however, are absent at the Fe(001) Fermi level. Even if they were present, the corresponding electron transmission would be completely filtered out by the MgO(001) \cite{Butler2001}. On the other hand, the $\Delta_5$ band in V is situated 0.5 eV above $E_F$ and it is also strongly attenuated by the MgO \cite{Butler2001}. Therefore, the low bias electron (spin) transport through V/MgO/Fe from the $\Delta_2$ states in Vanadium could be possible if $\Delta_2$ to $\Delta_1$ symmetry transformation takes place close to the V/MgO interface.
This peculiar property of the V band structure was previously explored by the insertion of thin Vanadium layers into hybrid Fe/V/Fe/MgO/Fe MTJs, which provided an increase of TMR up to $800000\%$ through the creation of QWSs for $\Delta_1$ electrons confined inside the soft Fe layer between the MgO and Vanadium \cite{Feng2009}.

Interestingly, our \textit{ab-initio} simulations disclose that the surface atomic layers at the V/MgO interface have $\Delta_1$ symmetry. Moreover, the numerical results also reveal that the vanadium surface states at the V/MgO interface provide a low bias (below 100 mV) peak in the density of states (DOS), in qualitative agreement with scanning tunneling spectroscopy measurements on V(001) \cite{Bischoff2001}. Most importantly, though, the simulations show a strong increase of the vanadium surface DOS peak with the inclusion of interfacial SOC and a relatively weak transformation under electric fields up to 1 V/Angstrom (see lower part in Figure \ref{Fig4}c). \farkhad{The effect of the SOC on the DOS is pronounced because the Fermi level without SOC lies within a gap for the $\Delta_1$ electrons (DOS from Figure \ref{Fig4}c is a histogram of the band along the $\Gamma-$H ($\Delta$) line, $\pm~0.4$ eV around $E_F$), which is quenched when the SOC is turned on because of the symmetry mixing ($\Delta_1$ gets scattered into $\Delta_{2}$ and vice versa, as shown in Figure \ref{Fig4}b), so the DOS gap gets filled with these states. There is also no $\Delta_5$ at $E_F$ without SOC, while with SOC $\Delta_5$ (not shown in the Figure) and $\Delta_1$ symmetries appear. Therefore, the enhancement of DOS when SOC is turned on mainly reflects the creation of $\Delta_1$ from $\Delta_{2}$.}

These results allow for a simple qualitative explanation of the main finding. As long as the switching of the device between the P and AP states redistributes the total voltage between the two parts (V/MgO/Fe and Fe/MgO/Fe/Co), the low bias conductance should increase faster with the total bias in the P state rather than in the AP state, due to the presence of the interfacial DOS peak providing the increase of TMR in the low bias regime.

Quantitatively, the TMR($V$) characteristic of the hybrid double barrier V/MgO/Fe/MgO/Fe/Co structure can be explained based on individual current-voltage characteristics of the bottom V/MgO/Fe and top Fe/MgO/Fe/Co MTJ subsystems considered as in series nonlinear devices (Figure \ref{Fig4}d inset). This decomposition in serial devices can be done because the 10 nm thickness of the middle Fe layer, larger than the coherence length for both the majority and minority spins, prevents coherent tunnelling across the two barriers. Therefore, the transport mechanisms in the double barrier N/F1/F2 junction is a sequential tunnelling in N/F1 and F1/F2 in series MTJs. However, the simulation of the serial device characteristics is not a simple voltage divider problem, because the voltage distribution on each component depends on the individual voltage-dependent nonlinear resistance. Therefore, the in-series circuit problem requires the resolution of a nonlinear circuit equations. This can be done in few steps, by considering the Kirchhoff voltage and current laws (KVL, KCL): (i) The KCL charge continuity condition, written as $I_1=I_2=I$ (where $I_1$ and $I_2$ are the currents passing through the N/F1 and F1/F2 barriers respectively), allows to determine the corresponding individual voltage drops $V_1$ and $V_{2(P/AP)}$ at each barrier from the individual i(V) characteristics of the N/F1 and F1/F2 nonlinear resistors. (ii) Then, the KVL will provide the total voltage drop on the serial device  $V=V_1+V_{2(P/AP)}$ for the current $I$ for which we determined the individual voltage drops. (iii) Finally, the serial circuit conductance would be $G_{P/AP}(V)=I/V$ and the corresponding $TMR(V)=[G_P(V)-G_{AP}(V)]/G_{AP}(V)$. This algorithm has been applied in our case considering as initial individual characteristics the experimental $G_2(V)$ in P and AP configurations (denoted as $G_{2(P/AP)}(V)$) for a F1/F2 MTJ, and a parametrized characteristic for the N/F1 junction issued from the fit of the experimental $G_1(V)$ of a N/F1 junction, adjusted to obtain the best fit of the final TMR(V) curve for the N/F1/F2 junction. The result of this simulation (Figure \ref{Fig4}(d)) explains the TMR variation with voltage in the hybrid double barrier MTJ in terms of competition between two main effects: (i) the standard decrease with voltage of the TMR in the top Fe/MgO/Fe/Co (F1/F2) MTJ (see Figure \ref{Fig2}) and (ii) the voltage induced increase in conductivity in the bottom V/MgO/Fe (N/F1) MTJ.  

\farkhad{The relatively low TMR at small biases is mainly a consequence of the low conductance of the Fe/MgO/V tunnel barrier (see Figure \ref{Fig1}), which provides an additional fixed resistance to the junction, therefore reducing the relative resistance difference between the P and AP states (i.e. the TMR). On the other hand, the same Fe/MgO/V interface is directly responsible for the enhancement of the TMR with applied bias. It’s characteristic $G(V)$ behavior allows the P and AP conductances to diverge, increasing the TMR up to $\sim0.5$ V where it becomes higher than for similar single barrier MTJs (Figure \ref{Fig2}), compensating the drawback at low bias with an advantage at intermediate and high voltages. At the same time, the fixed resistance of the Fe/MgO/V interface will increase the resistance-area product (RA) of the structure, which is another important parameter for applications. To the best of our knowledge, there isn’t any MTJ structure that allows for both an increasing of the TMR with bias and a low RA, although thinner tunnel barriers may allow for qualitatively similar behavior with a lower RA (calculations are, in fact, done with thinner barriers since the 2-3 nm of MgO are computationally prohibitive to simulate). While the reduction of the RA factor is important for applications in memory elements, it’s not so crucial for magnetic sensors where the increasing of TMR with voltage could be a bigger advantage.}

\section{Conclusion}
In summary, we introduce conceptually new hybrid magnetic tunnel junctions, consisting of standard MTJs sequentially coupled to a strongly nonlinear part, where the conductance is mainly provided by SOC-controlled interfacial states with symmetries different from the bulk states. This configuration provides a robust increase of the TMR with bias up to 0.5 V in a wide temperature range and unprecedented for room temperature spintronics high output voltage values. Our approach demonstrates the importance of the electron-symmetry-protected surface states in metallic interfaces, and is expected to push the applicability range of spintronic devices towards higher biases.

\section{Methods}
\subsection{Samples growth and characterization} 

The MTJ multilayer stacks have been grown by molecular beam epitaxy (MBE) in a chamber with a base pressure of $5\times10^{-11}$ mbar following the procedure described in Ref.\cite{Tiusan2007}. The samples were grown on (001) MgO substrates. Then, a 10 nm thick seed of anti-diffusion MgO underlayer is grown on the substrate to trap the C from it before the deposition of the Fe (or V). Then, the MgO insulating layer is epitaxially grown by e-beam evaporation, the thickness being approximately $\sim2$ nm, and so on with the rest of the layers. Each layer is annealed at 450 ºC for 20 mins for flattening. After the MBE growth, all the MTJ multilayer stacks are patterned in micrometre-sized square junctions by UV lithography and Ar ion etching, controlled step-by-step \textit{in situ} by Auger spectroscopy.

\subsection{Experimental measurement methods}

The measurements are performed using room temperature \cite{Herranz2010} and low temperature \cite{Martinez2018,Martinez2020} setups. The first consists of a sample holder inside a small low vacuum chamber, which is connected to a Keithley 2700 current source and a DMM-322 PCI voltimeter card, both controlled by a computer. The magnetic field used to switch the magnetic states is applied using a solenoid connected to a KEPCO 10-100 current/voltage source., which is also controlled with the computer.  Low temperature measurements are based on a JANIS$^{\tiny{\textregistered}}$ He$^3$ cryostat (the minimum attainable temperature is 0.3 K). The magnetic field is varied using a 3D vector magnet consisting of one solenoid (Z axis) with $H_{max}=3.5$ T and two Helmholtz coils (X and Y axis) with $H_{max}=1$ T. The temperature is measured and controlled using a LakeShore 340 controller. The current is applied with a Keithley K220 current source, and the voltage is measured with a DMM-552 PCI voltimeter card.

In both experimental configurations, the different magnetic states are obtained as follows: first, the field is set to $H=1000$ Oe (in the in-plane direction of the sample) saturating the magnetization of the two ferromagnetic layers, and then going back to zero field obtaining the P state. The AP state is obtained by applying a small magnetic field in the opposite direction, not exceeding the hard ferromagnetic layer coercive field of about 500 Oe.

\subsection{Ab-initio calculations methods}
\farkhad{The electronic structure calculations have been performed within a Full-Potential-Linear-Augmented-Plane-Wave method provided by the Wien2k code \cite{Wien2k}. A supercell model has been used to describe the surface features. The symmetry dependent densities of states are calculated as histograms from the dispersion bands $E(k)$ along the corresponding high symmetry directions (e.g. $\Gamma-\text{H}=\Delta$). The results are in full agreement with LKKR calculations from Ref. \cite{Feng2009}.} 

\medskip

\medskip
\textbf{Acknowledgements}

The work in Madrid was supported by Spanish Ministry of Science and Innovation (RTI2018-095303-B-C55) and Consejer\'ia de Educaci\'on e Investigaci\'on de la Comunidad de Madrid (NANOMAGCOST-CM Ref. P2018/NMT-4321) grants. F.G.A. acknowledges financial support from the Spanish Ministry of Science and Innovation, through the Mar\'ia de Maeztu Programme for Units of Excellence in R\&D (CEX2018-000805-M) and ``Accion financiada por la Comunidad de Madrid en el marco del convenio plurianual con la Universidad Autonoma de Madrid en Linea 3: Excelencia para el Profesorado Universitario''. C.T. acknowledges funding from the project ``MODESKY'' ID PN-III-P4-ID-PCE-2020-0230, No. UEFISCDI:PCE 4/04.01.2021


\medskip


\begin{thebibliography}{99}

\bibitem{Butler2001} W. H. Butler, X.-G. Zhang, T. C. Schulthess and J. M. MacLaren, Spin-dependent tunneling conductance of Fe/MgO/Fe sandwiches, \href{https://journals.aps.org/prb/abstract/10.1103/PhysRevB.63.054416}{Phys. Rev. B \textbf{63}, 054416 (2001).}

\bibitem{Mathon2001} J. Mathon and A. Umerski, Theory of tunneling magnetoresistance of an epitaxial Fe/MgO/Fe(001) junction, \href{https://journals.aps.org/prb/abstract/10.1103/PhysRevB.63.220403}{Phys. Rev. B \textbf{63}, 220403(R) (2001).}

\bibitem{Ney2003} A. Ney, C. Pampuch, R. Koch and K. H. Ploog, Programmable computing with a single magnetoresistive element, \href{https://www.nature.com/articles/nature02014}{Nature \textbf{425}, 485 (2003).}

\bibitem{Lenz2006} J. Lenz and S. Edelstein, Magnetic sensors and their applications, \href{https://ieeexplore.ieee.org/document/1634415}{IEEE Sensors Journal \textbf{6}, 3, 631 (2006).}

\bibitem{Parkin2004} S. S. P. Parkin, C. Kaiser, A. Panchula, P. M. Rice, B. Hughes, M. Samant and S.-H. Yang, Giant tunnelling magnetoresistance at room temperature with MgO (100) tunnel barriers, \href{https://www.nature.com/articles/nmat1256}{Nat. Mater. \textbf{3}, 862 (2004).}

\bibitem{Yuasa2004} S. Yuasa, T. Nagahama, A. Fukushima, Y. Suzuki and K. Ando, Giant room temperature magnetoresistance in single-crystal Fe/MgO/Fe magnetic tunnel junctions,  \href{https://www.nature.com/articles/nmat1257/}{Nat. Mater. \textbf{3}, 868 (2004).}

\bibitem{Waldron2006} D. Waldron, V. Timoshevskii, Y. Hu, K. Xia and H. Guo,  First Principles Modeling of Tunnel Magnetoresistance of Fe/MgO/Fe Trilayers, \href{https://journals.aps.org/prl/abstract/10.1103/PhysRevLett.97.226802}{Phys. Rev. Lett. \textbf{97}, 226802 (2006).}

\bibitem{Guerrero2007} R. Guerrero, D. Herranz and F. G. Aliev, High bias voltage effect on spin-dependent conductivity and shot noise in carbon-doped Fe(001)/MgO(001)/Fe(001) magnetic tunnel junctions, \href{https://aip.scitation.org/doi/10.1063/1.2793619}{Appl. Phys. Lett. \textbf{91}, 132504 (2007).}

\bibitem{Herranz2010} D. Herranz, F. Bonell, A. Gomez-Ibarlucea, S. Andrieu, F. Montaigne, R. Villar, C. Tiusan and F. G. Aliev, Strongly suppressed 1/f noise and enhanced magnetoresistance in epitaxial Fe–V/MgO/Fe magnetic tunnel junctions, \href{https://doi.org/10.1063/1.3430064}{Appl. Phys. Lett. \textbf{96}, 202501 (2010).} 

\bibitem{Parkin1997} \farkhad{S. Zhang, P. M. Levy, A. C. Marley and S. S. P. Parkin, Quenching of Magnetoresistance by Hot Electrons in Magnetic Tunnel Junctions, \href{https://journals.aps.org/prl/abstract/10.1103/PhysRevLett.79.3744}{Phys. Rev. Lett. \textbf{79}, 3744 (1997).}}

\bibitem{Moodera1995} \farkhad{J. S. Moodera, L. R. Kinder, T. M. Wong and R. Meservey, Large Magnetoresistance at Room Temperature in Ferromagnetic Thin Film Tunnel Junctions, \href{https://journals.aps.org/prl/abstract/10.1103/PhysRevLett.74.3273}{Phys. Rev. Lett. \textbf{74}, 3273 (1995).}}

\bibitem{Zhang2004} C. Zhang, X.-G. Zhang, P. S. Krsti\'{c}, H. Cheng, W. H. Butler and J. M. MacLaren, Electronic structure and spin-dependent tunneling conductance under a finite bias, \href{https://journals.aps.org/prb/abstract/10.1103/PhysRevB.69.134406}{Phys. Rev. B \textbf{69}, 134406 (2004).}

\bibitem{Nagahama2007} T. Nagahama, T. S. Santos and J. S. Moodera, Enhanced Magnetotransport at High Bias in Quasimagnetic Tunnel Junctions with EuS Spin-Filter Barriers, \href{https://journals.aps.org/prl/abstract/10.1103/PhysRevLett.99.016602}{Phys. Rev. Lett. \textbf{99}, 016602 (2007).}

\bibitem{Lu2005} Z.-Y. Lu, X.-G. Zhang and S. T. Pantelides, Spin-Dependent Resonant Tunneling through Quantum-Well States in Magnetic Metallic Thin Films, \href{https://journals.aps.org/prl/abstract/10.1103/PhysRevLett.94.207210}{Phys. Rev. Lett. \textbf{94}, 207210 (2005).}

\bibitem{PeraltaRamos2008}J. Peralta-Ramos, A. M. Llois, I. Rungger and S. Sanvito, I-V curves of Fe/MgO (001) single- and double-barrier tunnel junctions, \href{https://journals.aps.org/prb/abstract/10.1103/PhysRevB.78.024430}{Phys. Rev. B \textbf{78}, 024430 (2008).}

\bibitem{Xiang2019} Q. Xiang, H. Sukegawa, M. Belmoubarik, M. Al‐Mahdawi, T. Scheike, S. Kasai, Y. Miura and S. Mitani, Realizing Room‐Temperature Resonant Tunnel Magnetoresistance in Cr/Fe/MgAl$_2$O$_4$ Quasi‐Quantum Well Structures, \href{https://onlinelibrary.wiley.com/doi/full/10.1002/advs.201901438}{Advanced Science 6, 20 (2019).}

\bibitem{Niizeki2008} T. Niizeki, N. Tezuka and K. Inomata, Enhanced Tunnel Magnetoresistance due to Spin Dependent Quantum Well Resonance in Specific Symmetry States of an Ultrathin Ferromagnetic Electrode, \href{https://journals.aps.org/prl/abstract/10.1103/PhysRevLett.100.047207}{Phys. Rev. Lett. \textbf{100}, 047207 (2008).}

\bibitem{Kowalska2019} E. Kowalska, A. Fukushima, V. Sluka, C. Fowley, A. K\'akay, Y. Aleksandrov, J. Lindner, J. Fassbender, S. Yuasa and A. M. Deac, Tunnel magnetoresistance angular and bias dependence enabling tuneable wireless communication, \href{https://www.nature.com/articles/s41598-019-45984-5}{Sci. Rep. \textbf{9}, 9541 (2019).}

\bibitem{Montaigne1998} F. Montaigne, J. Nassar, A. Vaurès, F. Nguyen Van Dau, F. Petroff, A. Schuhl and A. Fert, Enhanced tunnel magnetoresistance at high bias voltage in double-barrier planar junctions, \href{hhttps://aip.scitation.org/doi/abs/10.1063/1.122604}{Appl. Phys. Lett. 73, 2829 (1998).}

\bibitem{Tiusan2006} C. Tiusan, F. Greullet, M. Sicot, M. Hehn, C. Bellouard, F. Montaigne, S. Andrieu and A. Schuhl, Engineering of spin filtering in double epitaxial tunnel junctions, \href{https://aip.scitation.org/doi/10.1063/1.2166592}{Journal of Applied Physics \textbf{99}, 08A903 (2006).}

\bibitem{Gan2010} H. D. Gan, S. Ikeda, W. Shiga, J. Hayakawa, K. Miura, H. Yamamoto, H. Hasegawa, F. Matsukura, T. Ohkubo, K. Hono and H. Ohno, Tunnel magnetoresistance properties and film structures of double MgO barrier magnetic tunnel junctions, \href{http://dx.doi.org/10.1063/1.3429594}{Appl. Phys. Lett. \textbf{96}, 192507 (2010).}

\bibitem{Tezuka2016} N. Tezuka, S. Oikawa, I. Abe, M. Matsuura, S. Sugimoto, K. Nishimura and T. Seino, Perpendicular Magnetic Tunnel Junctions With Low Resistance-Area Product: High Output Voltage and Bias Dependence of Magnetoresistance, \href{https://ieeexplore.ieee.org/document/7498675}{IEEE Magnetics Letters, \textbf{7}, 3104204 (2016).}

\bibitem{Useinov2012}A. Useinov, O. Mryasov and J. Kosel, Output voltage calculations in double barrier magnetic tunnel junctions with asymmetric voltage behavior, \href{https://doi.org/10.1016/j.jmmm.2012.04.025}{Journal of Magnetism and Magnetic Materials \textbf{324}, 18, 2844-2848 (2012).}

\bibitem{Mukaiyama2016} K. Mukaiyama, S. Kasai, Y. K. Takahashi, K. Kondou, Y. Otani, S. Mitani and K. Hono, High output voltage of magnetic tunnel junctions with a Cu(In$_{0.8}$Ga$_{0.2}$)Se$_2$ semiconducting barrier with a low resistance-area product, \href{https://doi.org/10.7567/APEX.10.013008}{Applied Physics Express, \textbf{10}, 1 (2016).}

\bibitem{Tiusan2006_2} C. Tiusan, M. Sicot, M. Hehn, C. Belouard, S. Andrieu, F. Montaigne and A. Schuhl, Fe/MgO interface engineering for high-output-voltage device applications \href{https://aip.scitation.org/doi/10.1063/1.2172717}{Appl. Phys. Lett. \textbf{88}, 062512 (2006).}

\bibitem{Feng2009} X. Feng, O. Bengone, M. Alouani, I. Rungger and S. Sanvito, Interface and transport properties of Fe/V/MgO/Fe and Fe/V/Fe/MgO/Fe magnetic tunneling junctions, \href{https://journals.aps.org/prb/abstract/10.1103/PhysRevB.79.214432}{Phys. Rev. B \textbf{79}, 214432 (2009).}

\bibitem{Bischoff2001} M. M. J. Bischoff, C. Konvicka, A. J. Quinn, M. Schmid, J. Redinger, R. Podloucky, P. Varga and H. van Kempen, Influence of Impurities on Localized Transition Metal Surface States: Scanning Tunneling Spectroscopy on V(001), \href{https://journals.aps.org/prl/abstract/10.1103/PhysRevLett.86.2396}{Phys. Rev. Lett. \textbf{86}, 2396 (2001)}

\bibitem{Tiusan2007} C. Tiusan, M. Hehn, F. Montaigne, F. Greullet, S. Andrieu and A. Schuhl, Spin tunneling phenomena in single crystal magnetic tunnel junction systems, \href{https://iopscience.iop.org/article/10.1088/0953-8984/19/16/165201/meta}{J. Phys.: Condens. Matter \textbf{19}, 165201 (2007).}

\bibitem{Martinez2018} I. Mart\'inez, C. Tiusan, M. Hehn, M. Chshiev and F. G. Aliev, Symmetry Broken Spin Reorientation Transition in Epitaxial MgO/Fe/MgO Layers with Competing Anisotropies, \href{https://www.nature.com/articles/s41598-018-27720-7}{Sci. Rep. \textbf{8}, 9463 (2018).}

\bibitem{Martinez2020} I. Mart\'inez, P. H\"ogl, C. Gonz\'alez-Ruano, J. P. Cascales, C. Tiusan, Y. Lu, M. Hehn, A. Matos-Abiague, J. Fabian, I. Zutic and F. G. Aliev, Interfacial spin-orbit coupling: a platform for superconducting spintronics,  \href{https://journals.aps.org/prapplied/abstract/10.1103/PhysRevApplied.13.014030}{Phys. Rev. Appl. \textbf{13}, 014030 (2020).}

\bibitem{Wien2k} \farkhad{P. Blaha, K.Schwarz, F. Tran, R. Laskowski, G.K.H. Madsen and L.D. Marks, WIEN2k: An APW+lo program for calculating the properties of solids, \href{https://aip.scitation.org/doi/full/10.1063/1.5143061}{J. Chem. Phys. \textbf{152}, 074101 (2020).}}


\end{thebibliography}
\end{document}